\documentclass[prd,nofootinbib,preprint,showpacs]{revtex4}

\usepackage{amsmath}

\def\lqcd{\Lambda_\text{QCD}}

\begin{document}
\preprint{\vbox{\hbox{SCIPP-02/19} \hbox{WIS/40/02-Sep-DPP}}}

\title{Time Variations in the Scale of Grand Unification}

\author{Michael Dine}
\affiliation{Santa Cruz Institute for Particle Physics\\
     Santa Cruz CA 95064, USA\\ 
   }

\author{Yosef Nir}
\affiliation{Particle Physics Department \\ Weizmann Institute of
  science \\ Rehovot 76100, Israel}

\author{Guy \surname{Raz}}
\affiliation{Particle Physics Department \\ Weizmann Institute of
  science \\ Rehovot 76100, Israel}

\author{Tomer \surname{Volansky}}
\affiliation{Particle Physics Department \\ Weizmann Institute of
  science \\ Rehovot 76100, Israel}


\begin{abstract}
  We study the consequences of time variations in the scale of grand
  unification, $M_{\rm U}$, when the Planck scale and the value of the
  unified coupling at the Planck scale are held fixed. We show that
  the relation between the variations of the low energy gauge
  couplings is highly model dependent. It is even possible, in
  principle, that the
  electromagnetic coupling $\alpha$ varies, but the strong coupling
  $\alpha_3$ does not (to leading approximation). We investigate whether
  the interpretation of recent observations of quasar absorption lines
  in terms of time variation in $\alpha$ can be accounted for by time
  variation in $M_{\rm U}$. Our formalism can be applied to any
  scenario where a time variation in an intermediate scale induces,
  through threshold corrections, time variations in the effective low
  scale couplings.
\end{abstract}

\maketitle

\section{Introduction}
\label{sec:introduction}
Recent observations of quasar absorption lines have been interpreted
as indicating variation in the fine structure constant on cosmological
time scales \cite{Webb:2000mn,Murphy:2002jx},
\begin{equation}\label{tivaal}
  {\delta\alpha\over\alpha}=(-0.57\pm0.10)\times10^{-5}\ {\rm at}\
  z\approx0.2-3.7.
\end{equation}
In the context of grand unified theories (GUTs), such a time variation
in $\alpha$ may be related to time variations in other gauge couplings
and, in particular, to variations in the strong coupling constant
$\alpha_3$ or, equivalently, in the QCD scale $\lqcd$. Such related
variations affect various observables in interesting ways
\cite{Calmet:2001nu,Langacker:2001td,Dent:2001ga,Flambaum:2002de,Ichikawa:2002bt,Chacko:2002mf,Calmet:2002ja,Olive:2002tz}.

Most previous works have assumed that the variation in $\alpha$ is
induced by a variation in the unified coupling at the Planck scale,
$\alpha_{\rm U}(M_{\rm Pl})$
\cite{Calmet:2001nu,Langacker:2001td,Dent:2001ga,Flambaum:2002de,Ichikawa:2002bt,Olive:2002tz}.
The motivation for such studies comes mainly from string theories,
where the value of $\alpha_{\rm U}(M_{\rm Pl})$ is determined by the value
of the dilaton and other fields. We are here interested in another
possibility, 
namely that the variation in $\alpha$ is induced by a variation in the
GUT scale $M_{\rm U}$, with $M_{\rm Pl}$ and $\alpha_{\rm U}(M_{\rm
Pl})$ held fixed.  In ref. \cite{Dine:2002se}, it is argued that if
grand unification (in the sense of unification of the gauge
interactions in a four-dimensional gauge group) arises within
string theory, the GUT scale is likely to be a modulus.\footnote{Of 
  course, in string theory, coupling unification does not necessarily
  require a unified gauge group.}
Such a situation can also arise in the framework of field
theories, where symmetry breaking scales are determined by vacuum
expectation values (VEVs) of scalar fields, but coupling constants are
not.  In either case, understanding why the modulus has
the requisite properties is a difficult problem, which
we will not address.  We will show that the relation between
$\delta\alpha/\alpha$ and $\delta\lqcd/\lqcd$ is highly model
dependent. In particular, in contrast to the case that both are
induced by $\delta\alpha_{\rm U}/\alpha_{\rm U}\neq0$, it may happen
that $\delta\lqcd/\lqcd\ll\delta\alpha/\alpha$ (though this is not the
generic case).

We are aware of two previous works that are closely related to
ours. In ref. \cite{Calmet:2002ja}, the consequences of variations
of $M_{\rm U}$ have been analyzed. It was assumed, however, that
$\alpha_{\rm U}(M_{\rm U})$ remains fixed. It is then necessary that
$\alpha_{\rm U}(M_{\rm Pl})$ also varies in a correlated way. We think
that such a scenario is less plausible than the one that we investigate.
In ref. \cite{Chacko:2002mf}, the consequences for the low energy
couplings of variations in the mass of
particles have been analyzed. Conceptually, our work takes a similar
direction. In their discussion of
grand unified theories, ref. \cite{Chacko:2002mf} examine a set of models in which
only the masses of a subset of GUT mass particles vary in time; in particular,
only the masses of
color-neutral, charged particles shift. In contrast, our basic assumption is
that all heavy masses (of GUT gauge bosons, fermions and scalars) are
proportional to a single breaking scale, $M_{\rm U}$, and it is the
variation of this scale that we study.\footnote{To understand the
difference, one may think of the consequences of a variation in the VEV of the
single Higgs doublet of the Standard Model, which would shift all
masses, or a shift in the VEV of a hypothetical Higgs triplet, which
would shift only neutrino masses.}

In this work we focus on the resulting relations between the low
energy couplings. We do not concern ourselves here with the origin of
the scale variation, nor with the connection to higher scale
physics. Discussions of these topics can be found
in~\cite{Campbell:1994bf,Chacko:2002mf}. In addition, there are
fundamental issues we will also not address. As pointed out in
\cite{Banks:2001qc}, it is difficult to understand, in the
framework of local quantum field theory, how couplings could vary by so
large an amount, without an enormous variation in the vacuum energy.
This problem is not significantly different if it is changes in the
unification scale which are responsible for this variation than if the
coupling at the Planck or unification scale changes.

The plan of this paper is as follows. In section~\ref{sec:formalizm}
we present the formalism for the variation of low energy couplings
induced by variations in the scale of threshold corrections. In
section~\ref{sec:observables} we obtain the relations between the
variation of relevant cosmological observables and argue that, within
our framework, they depend on a single parameter. In
section~\ref{sec:test-against-exper} we apply the formalism to the
experimental data. We use it to constrain the parameter of
section~\ref{sec:formalizm}.  In section~\ref{sec:impl-vari-gut} we
test specific models of grand unified theories with this constraint.
We show that a variation in the GUT scale is unlikely to explain the
claimed variation in $\delta\alpha/\alpha$.  In
section~\ref{sec:generalizations} we explain how our formalism and
results apply much more generically, to any model where there is a
time variation in a scale where threshold corrections take place. We
also mention some further subleading effects which might be
non-negligible.  We conclude in section~\ref{sec:conclusions}.

\section{Formalism}
\label{sec:formalizm}
We assume that the variation in couplings is due to a variation in
a single intermediate scale, $M_{\rm U}$, where GUT threshold
corrections take place. The gauge couplings $\alpha_i(Q)$ at a scale
$Q< M_{\rm U}$ are given to one loop by the RGE solution
\begin{equation}
  \label{eq:1}
  \alpha^{-1}_i(Q)=\alpha^{-1}_{\rm U}(M_{\rm U}) +
  \frac{b_i}{2\pi}\ln\left(\frac{Q}{M_{\rm U}}\right),
\end{equation}
where $\alpha_{\rm U}(M_{\rm U})$ is the unified coupling at the scale
$M_{\rm U}$ and $b_i$ is the beta function coefficient (between
$M_{\rm U}$ and $Q$) for $\alpha_i$.
The scale $M_{\rm U}$ is related to the mass scale of heavy degrees of
freedom that are integrated out, particularly the heavy GUT gauge
bosons. Since the particle content of the theory below and above
$M_{\rm U}$ is different, the RGE coefficient $b_i$ changes at $M_{\rm
  U}$. In particular, one obtains
\begin{equation}
  \label{eq:2}
  \alpha^{-1}_{\rm U}(M_{\rm U})=\alpha^{-1}_{\rm U}(M_{\rm Pl}) +
  \frac{b_{\rm U}}{2\pi}\ln\left(\frac{M_{\rm U}}{M_{\rm Pl}}\right),
\end{equation}
where $b_{\rm U}$ is the beta function coefficient for the unified
group.
Combining~\eqref{eq:1} and~\eqref{eq:2} we write
\begin{equation}
  \label{eq:3}
  \alpha^{-1}_i(Q)=\alpha^{-1}_{\rm U}(M_{\rm Pl}) +
  \frac{b_{\rm U}}{2\pi}\ln\left(\frac{M_{\rm U}}{M_{\rm Pl}}\right)+
  \frac{b_i}{2\pi}\ln\left(\frac{Q}{M_{\rm U}}\right).
\end{equation}

Now we allow the GUT scale, $M_{\rm U}$, to change by an amount of
$\delta M_{\rm U}$, while holding $\alpha_{\rm U}(M_{\rm Pl})$ and
$M_{\rm Pl}$ fixed. The resulting variation in the low scale couplings
can be calculated from~\eqref{eq:3} (a similar derivation, for
$\alpha_\text{EM}$, appears in~\cite{Chacko:2002mf}):
\begin{equation}
  \label{eq:4}
  \frac{\delta\alpha_i(Q)}{\alpha_i(Q)} = \alpha_i(Q)\frac{\Delta
  b_i}{2\pi}\frac{\delta M_{\rm U}}{M_{\rm U}}\;,
\end{equation}
where
\begin{equation}
  \label{eq:5}
  \Delta b_i \equiv b_{\rm U} - b_i\;.
\end{equation}

Using~\eqref{eq:4}, we can now relate the variation in low scale
parameters to the variation $\delta M_{\rm U}$. We focus on two parameters,
the electromagnetic coupling $\alpha$ and the QCD scale $\lqcd$:
\begin{equation}
  \label{eq:6}
  \alpha^{-1}=\frac{5}{3}\alpha^{-1}_1(0)+\alpha^{-1}_2(0)\;,
\end{equation}
\begin{equation}
  \label{eq:7}
  \lqcd = M_Z^{23/27}m_b^{2/27}m_c^{2/27}\exp\left(-\frac{2\pi}{9\,\alpha_3(M_Z)}\right)\;.
\end{equation}
Neglecting effects of threshold corrections below $M_{\rm U}$ (which
are discussed later), we obtain the relations:
\begin{equation}
  \label{eq:8}
  \frac{\delta\alpha}{\alpha}=\frac{\alpha}{2\pi}\left(\frac{5}{3}\Delta
  b_1+\Delta b_2\right)\ \frac{\delta M_{\rm U}}{M_{\rm U}}\;,
\end{equation}
\begin{equation}
  \label{eq:9}
  \frac{\delta\lqcd}{\lqcd}=\frac{\Delta b_3}{9}\
  \frac{\delta M_{\rm U}}{M_{\rm U}}.
\end{equation}

Finally, we use~\eqref{eq:8} and~\eqref{eq:9} to relate the variation
of the two parameters:
\begin{eqnarray}
  \label{eq:10}
    \frac{\delta\lqcd}{\lqcd} &=& \frac{2\pi}{9\alpha}\
    \frac{\Delta b_3}{\frac{5}{3}\Delta b_1+\Delta b_2}\
    \frac{\delta\alpha}{\alpha}\nonumber\\
    &\equiv& R\ \frac{\delta\alpha}{\alpha}.
  \end{eqnarray}
The factor $R\approx\frac{95.7\Delta b_3}{\frac{5}{3}\Delta b_1+\Delta b_2}$
plays a crucial role in our
analysis.  As we show later, its value is highly GUT-model dependent.

\section{Observables}
\label{sec:observables}
In this section, we use~\eqref{eq:10} to obtain expressions for the
variation of several cosmological observables. The primordial $^4$He
abundance is given by~\cite{Kolb:1990vq,Bergstrom:1999wm}
\begin{equation}
  \label{eq:13}
  Y_4 \equiv \frac{2(n_n/n_p)_\text{NS}}{1+(n_n/n_p)_\text{NS}}\;,
\end{equation}
where the ratio of the number density of the neutron to that of the
proton at nucleosynthesis is given by
\begin{equation}
  \label{eq:14}
  (n_n/n_p)_\text{NS}\simeq 0.8\times e^{-Q/T_D} \;.
\end{equation}
Here $T_D\simeq0.8$ MeV is the decoupling temperature and $Q$ is the
mass difference between the neutron and the proton. Since $T_D$ is a
function of $M_{\rm Pl}$ and $G_F$, it is independent of the Standard
Model gauge couplings and we therefore neglect its variation. The mass
difference $Q$, on the other hand, depends on both $\alpha$ and
$\lqcd$~\cite{Gasser:1982ap}:
\begin{equation}
  \label{eq:15}
  Q \simeq 2.05\ \text{MeV}+C\alpha\lqcd \;,
\end{equation}
where $C$ is a dimensionless order one parameter. Assuming variations
in $\alpha$ and $\lqcd$, the value of $Q$ is shifted [eq. \eqref{eq:15}]
from its present-day value of $Q\simeq1.29\text{ MeV}$, and
consequently $(n_n/n_p)_\text{NS}$ would be shifted [eq. \eqref{eq:14}]
from its `standard' value of $(n_n/n_p)_\text{NS}\simeq \frac{1}{7}$.
We thus obtain the change in $Y_4$ compared to its value
calculated with the present values of the coupling constants:
\begin{equation}
  \label{eq:16}
  \begin{split}
    \frac{\delta Y_4}{Y_4} & = \frac{1}{1+(n_n/n_p)_\text{NS}}\,
    \frac{2.05\ \text{MeV}-Q}{T_D}\left(\frac{\delta\alpha}{\alpha} +
      \frac{\delta\lqcd}{\lqcd}\right) \\
    & \simeq 0.8\left(1+R\right)\frac{\delta\alpha}{\alpha}\;.
\end{split}
\end{equation}

Two other relevant observables are $X$ and $Y$. The $X$ parameter
gives the ratio between the hyperfine 21cm neutral hydrogen absorption
transition to an optical resonance transition~\cite{Cowie:1995sz}:
\begin{equation}
  \label{eq:18}
  X \equiv \alpha^2 g_p \left(\frac{m_e}{m_p}\right),
\end{equation}
where $m_e$ is the electron mass, $m_p$ is the proton mass, and $g_p$
is the proton gyromagnetic moment. The $Y$ parameter determines the
molecular hydrogen transition in the early
universe~\cite{Potekhin:1998mf,Ivanchik:2001ji},
\begin{equation}
  \label{eq:20}
  Y \equiv \frac{m_p}{m_e}\;.
\end{equation}
The proton mass is proportional, at first order, to
$\lqcd$~\cite{Langacker:2001td}:
\begin{equation}
  \label{eq:17}
  \frac{\delta m_p}{m_p} \simeq \frac{\delta\lqcd}{\lqcd} =
  R\frac{\delta\alpha}{\alpha}\;.
\end{equation}
Assuming that the variations of $m_e$ and
 $g_p$ are negligible, we obtain:
\begin{equation}
  \label{eq:19}
  \frac{\delta X}{X} \simeq 2\frac{\delta\alpha}{\alpha}-\frac{\delta
  m_p}{m_p} \simeq \left(2- R\right)\frac{\delta\alpha}{\alpha}\;,
\end{equation}
\begin{equation}
  \label{eq:21}
  \frac{\delta Y}{Y} \simeq R \frac{\delta\alpha}{\alpha}\;.
\end{equation}

\section{Constraining $R$ with Experimental Data}
\label{sec:test-against-exper}
Values for the observed variation of various observables were obtained
through cosmological and nuclear studies. Table~\ref{tab:constraints}
summarizes these values (the details can be found in the quoted
references). Using the formalism of the previous section, we relate
the allowed range for each observable in Table~\ref{tab:constraints}
to an (in general, $R$-dependent) allowed variation in $\alpha$. The
results are also given in Table~\ref{tab:constraints}.

\begin{table}[htbp]
  \caption{Cosmological and nuclear values for the variation of
    various observables and their implications for $\delta\alpha/\alpha$.}
  \label{tab:constraints}
  \begin{ruledtabular}
    \begin{tabular}{lccc}
      Constraints & Redshift $z$ & Reference & Implied constraint on
      $\delta\alpha/\alpha$ \\
      \hline
      $\displaystyle \frac{\delta Y_4}{Y_4} = (-6\pm7)\times 10^{-3}$ &
      $10^{10}$ & \cite{Bergstrom:1999wm,Avelino:2001nr} &
      $(-7\pm9)\times10^{-3}/(1+R)$ \medskip \\ 
      $\displaystyle \frac{\delta \alpha}{\alpha} = (-0.57 \pm 0.10)\times 10^{-5}$
      & $0.2\text{--}3.7$ & \cite{Webb:2000mn,Murphy:2002jx} & $
      (-0.57 \pm 0.10)\times 10^{-5}$ \medskip \\
      $\displaystyle \frac{\delta Y}{Y} = (5.7\pm 3.8)\times10^{-5}$ &
      $2.3 \text{--} 3$ & \cite{Ivanchik:2001ji} & $(5.7
    \pm 3.8)\times 10^{-5}/R$ \medskip \\
      $\displaystyle \frac{\delta X}{X} = (0.7\pm 1.1)\times10^{-5}$ &
      $1.8$ & \cite{Cowie:1995sz} & $(0.7
    \pm 1.1)\times 10^{-5}/(2-R)$ \medskip \\
      $\displaystyle \frac{\delta \alpha}{\alpha} = (-0.36 \pm
      1.44)\times 10^{-8}$ & $0.1$ & \cite{Damour:1996zw,Fujii:1998kn} &
      $(-0.36 \pm  1.44)\times 10^{-8}$
    \end{tabular}
  \end{ruledtabular}
\end{table}

For a given value of $R$, one obtains from Table~\ref{tab:constraints}
the implied variation of $\alpha$ at different $z$'s. The most
interesting result, however, arises from requiring consistency of the
three observables related to the intermediate redshift values.
Note that $\delta Y/Y$ and $\delta X/X$ are consistent with each other
for any value of $R$, because, at the $2\sigma$ level, they allow
$\delta\alpha/\alpha=0$, for which the value of $R$ becomes
irrelevant. The constraints become nontrivial when one considers the
claimed signal of $\delta\alpha/\alpha$ [eq. (\ref{tivaal})].
The requirement of consistency between the observed $\delta Y/Y$ at
$z\approx2.3\text{--}3$ and the observed $\delta \alpha/\alpha$ at
$z\approx 0.2\text{--}3.7$ favors a range in $R$ that is mostly
negative in sign and not very large in magnitude, $R={\cal
  O}(-10)$. The requirement of consistency between the observed
$\delta X/X$ at $z\approx1.8$ and the observed $\delta \alpha/\alpha$ at
$z\approx 0.2\text{--}3.7$ favors a range in $R$ that is mostly
positive in sign and, again, not very large in magnitude, $R={\cal
  O}(+3)$. The most interesting results follows from fitting
simultaneously all three observables. The result of a $\chi^2$
analysis is that the following range of $R$ is favored:
\begin{equation}
  \label{eq:26}
  -1 \lesssim R \lesssim +6 \;.
\end{equation}
Values of $R$ outside this range have a probability that is
$\lesssim5\%$ to yield the observed results. (Note that even the best
fit point, with $R\sim+2$, has only a probability $\sim15\%$ to yield
the observed results.)

This result can be understood qualitatively: When $|R|$ is large, a
value for $\delta\alpha/\alpha$ implies an even larger value for
$\delta\lqcd/\lqcd$. This larger variation dominates the shift of
nucleon masses. Since $Y$ is proportional to the proton mass while
$X$ is inversely propotional to it, their relative variations must be
of opposite signs, $\delta Y\delta X\leq 0$. The data, however,
indicate that their variations are both biased to be
positive. Consequently, only one of them can be consistent
with a non-zero variation of $\alpha$. (Which one depends on the sign
of $R$.) This situation is avoided only in case that the
variation in $\lqcd$ does not dominate $\delta X/X$ or, in other
words, in case of a small $R$.

To summarize, eq.~\eqref{eq:26} gives the consistent range for
the parameter $R$ in theories where the variation of the low energy
coupling constants is dominated by threshold effects at a single
varying scale.

\section{Implications for various GUT models}
\label{sec:impl-vari-gut}
In this section we apply our results to specific models. We focus
on the minimal supersymmetric standard model (MSSM) embedded in a
grand unified theory (GUT), in which the varying scale is the scale of
breaking of the unified group, $M_{\rm U}$. In general, this is the
mass scale of the GUT gauge supermultiplets, and we assume that it
characterizes also the masses of all the heavy chiral
supermultiplets. The one loop beta function coefficients below $M_{\rm
  U}$ are those of the MSSM:
\begin{equation}
  \label{eq:22}
  b_1= \frac{33}{5}\;,\qquad b_2=1\;,\qquad b_3=-3\;.
\end{equation}
Above the GUT scale, there is a single coefficient common to all gauge
couplings, $b_{\rm U}$. Thus, the ratio $R$ of eq. \eqref{eq:10} is
given by
\begin{equation}
  \label{eq:rgut}
  R={2\pi\over9\alpha}\ {b_{\rm U}+3\over\frac{8}{3}b_{\rm U}-12}\:.
\end{equation}

Before we discuss specific GUT models, let us make some general
observations:

1. For asymptotically large (positive or negative) values
of $b_{\rm U}$, we obtain
\begin{equation}\label{asymbu}
  |b_{\rm U}|\gg1\ \ \Longrightarrow\ \
   R\to\frac{\pi}{12\alpha}\simeq+36.
 \end{equation}
Large negative values of $b_{\rm U}$ are not relevant, as
demonstrated by the very conservative lower bounds set by examining
the contributions from the gauge supermultiplet and from the minimal
matter representations that are necessary to accommodate the MSSM
fields:
\begin{equation}\label{minbu}
  b_{\rm U}[SU(5)]>-8,\ \ \ b_{\rm U}[SO(10)]>-17.
\end{equation}
On the other hand, realistic models often have a large, positive
$b_{\rm U}$ and predict $R={\cal O}(40)$.

2. Negative values of $R$ are achieved only for a very limited
  range of $b_{\rm U}$:
  \begin{equation}\label{negbu}
    R<0\ {\rm for}\ -3<b_{\rm U}<+\frac{9}{2}.
    \end{equation}

3. For a hypothetical value of $b_{\rm U}=\frac{9}{2}$, $\alpha$
  does not vary (to the approximation that we use):
\begin{equation}\label{rinfty}
  b_{\rm U}=\frac{9}{2}\ \ \Longrightarrow\ \ |R|\to\infty.
 \end{equation}
This result explains the change of sign of $R$ between models with
$b_{\rm U}=4$ and $b_{\rm U}=5$, and the very large magnitude of $R$
in these models, $R\simeq-502(+574)$ for $b_{\rm U}=4(5)$.

4. For $b_{\rm U}=b_3=-3$, $\lqcd$
does not vary (to the approximation that we use):
\begin{equation}\label{rzero}
  b_{\rm U}=-3\ \ \Longrightarrow\ \ R=0.
\end{equation}
In general, the $\mathcal{O}(100)$ factor in eq.~\eqref{eq:10}, which
is due to the exponential dependence of $\lqcd$ on $\alpha_3$, leads
to a large $R$. In order to overcome this factor and obtain
$|R|\sim\mathcal{O}(1)$, the threshold correction of $\alpha_3$ needs
to be highly suppressed, {\it i.e.}, $\Delta b_3\lesssim 0.01\Delta
b_{1,2}$. For MSSM GUT models with varying $M_{\rm
  GUT}$, this would imply $-3.5 \lesssim b_{\rm U}\lesssim -2.5 $. In
particular, when $b_{\rm U}$ is shifted by one unit from the special
value of $-3$, one already obtains rather large values of $|R|$,
$R=+4.2(-5.5)$ for $b_{\rm U}=-4(-2)$. We learn that, if $R$ is to be
within the range of eq.~\eqref{eq:26}, the only acceptable integer
values for $b_{\rm U}$ are $-3$ and $-4$.

We now move on to discuss several examples of specific GUT models. The
$b_{\rm U}$ coefficients and the resulting $R$-factors of various
models are given in Table~\ref{tab:various_bs}. Each model is defined
by its representation content. (Note that, for example, $n_{\bf5}$
gives the number of SU(5) fundamentals plus the number of
antifundamentals.) All SU(5) models have the quark and lepton fields
in three generations of ${\bf10}+\overline{\bf5}$ and the Higgs fields
related to electroweak symmetry breaking in
${\bf5}+\overline{\bf5}$. The minimal model has, in addition, a single
${\bf24}$ to break $SU(5)\to G_{\rm SM}$. This model is, however,
excluded by combining constraints from coupling unification and from
proton decay \cite{Murayama:2001ur}. (It may still be viable with a
special flavor structure of the supersymmetric mixing matrices
\cite{Bajc:2002bv}.) A viable, though fine-tuned,
model can be constructed by adding another ${\bf5}+\overline{\bf5}$
pair \cite{Hisano:1992ne,Murayama:2001ur}. To naturally induce
doublet-triplet splitting, the ${\bf24}$ has to be replaced by, at
least, ${\bf50}+\overline{\bf50}+{\bf75}$
\cite{Masiero:1982fe,Grinstein:1982um,Altarelli:2000fu}.

All SO(10) models have the quark and lepton fields in three
generations of ${\bf16}$. The minimal Higgs sector has ${\bf
  45}+{\bf16}+\overline{\bf16}$ to break $SO(10)\to G_{\rm SM}$ and a
single ${\bf10}$ for electroweak symmetry breaking
\cite{Babu:1998wi}. Note that we assume here that the
breaking of $SO(10)\to G_{\rm SM}$ is done in one step, and that the
VEVs of all GUT-breaking Higgs fields vary together. The investigation
of a two-step breaking, $SO(10)\to SU(5)\to G_{\rm SM}$, with
independent variation of each scale, will be presented in future
work.\footnote{In case that the scale of $SO(10)\to SU(5)$ breaking
  varies but the scale of $SU(5)\to G_{\rm SM}$ breaking does not, the
result is effectively a variation in the unified coupling constant at
the $SU(5)$ breaking scale. This result shows how our formalism
can be applied also to models of time variation in $\alpha_{\rm U}$:
we should take $\Delta b_1=\Delta b_2=\Delta b_3$, which gives
$R\simeq36$.} Models where doublet-triplet splitting is achieved naturally
require an additional ${\bf10}$ and either an additional
${\bf16}+\overline{\bf16}$ pair \cite{Barr:1997hq} or a more
complicated Higgs sector that includes several ${\bf45}$ and ${\bf54}$
multiplets \cite{Babu:1993we}. In each of these models, one could
have a ${\bf126}+\overline{\bf126}$ pair in place of a
${\bf16}+\overline{\bf16}$ pair. Such a replacement increases $b_{\rm
  U}$ by $66$ and gives $R={\cal O}(40)$.

\begin{table}[htbp]
  \caption{Various GUT models, their beta function
    coefficient $b_{\rm U}$ and the resulting factor $R$.}
  \label{tab:various_bs}
  \begin{ruledtabular}
    \begin{tabular}{llllllrr}
      SU(5) & $n_{\bf 5}$ & $n_{\bf 10}$ & $n_{\bf 24}$
      & $n_{\bf 50}$ & $n_{\bf 75}$ & $b_{\rm U}$ & $R$ \\
      \hline
      & $5$ & $3$  & $1$ &    &     & $-3$ & $0$ \\
      & $7$ & $3$  & $1$ &    &     & $-2$ & $-5.5$ \\
      & $5$ & $3$  &     & $2$& $1$ & $+52$& $+41.5$ \\
       \hline
       SO(10) &   $n_{\bf 10}$ & $n_{\bf 16}$ & $n_{\bf 45}$
      & $n_{\bf 54}$ && $b_{\rm U}$ & $R$ \\
      \hline
     & $1$ & $5$  & $1$ &    && $-5$ & $+7.6$ \\
     & $2$ & $7$  & $1$ &    && $0 $ & $-23.9$ \\
     & $2$ & $5$  & $3$ & $2$&& $+36$& $+44.4$
     \end{tabular}
   \end{ruledtabular}
  \end{table}

Other models of interest are reviewed in, for example,
ref.~\cite{Mohapatra:1999vv}
and lead to similar results. We conclude that most GUT models give
$|R|\gg 1$. A remarkable exception is given by models with $b_{\rm
  U}=-3$. This possibility is realized in the minimal $SU(5)$ model
which, without a very special flavor structure, is, however,
phenomenologically excluded. Other models with
$b_{\rm U}=-3$ can be constructed within the framework of larger
groups. There is, however, little motivation to consider such a
scenario (which would require fine-tuning for the doublet-triplet
splitting) for the sole purpose of deriving the desired $b_{\rm
  U}$. We conclude that, within the framework of realistic GUT models
that avoid fine-tuning, our mechanism predicts $|R|\gg 1$.

When this result is confronted with eq.~\eqref{eq:26}, one is led to
the conclusion that supersymmetric GUT models with a varying GUT scale
are unlikely candidates to explain a variation in $\alpha$ of the size
given in eq.~\eqref{tivaal}.

\section{Generalization and Limitations}
\label{sec:generalizations}
Our formalism and many of our results have a much broader
applicapibility than the $M_{\rm U}$-variation in GUT models that we
have focussed on. The formalism applies to any
variation in a single intermediate scale where threshold corrections
take place, $\mu_{\rm th}$. Suppose that the couplings are held fixed
at some fixed high energy scale $\mu_0$. Define $b_i$ to be the one-loop
beta function coefficients between  $\mu_{\rm th}$ and a low scale $Q$,
while $b^\prime_i$ are the corresponding coefficients between  $\mu_0$
and $\mu_{\rm th}$. We now define, as a generalization of
eq. \eqref{eq:5},
\begin{equation}\label{delbgen}
  \Delta b_i\equiv b^\prime_i-b_i.
\end{equation}
With this definition, eq. \eqref{eq:8} for $\delta\alpha/\alpha$,
eq. \eqref{eq:9} for $\delta\lqcd/\lqcd$, and eq. \eqref{eq:10} for
the ratio $R$, still apply.\footnote{One has to be careful about the
  normalization of $\alpha_1$ in non-GUT models. For example, if we
  take the conventional definition of $U(1)_Y$ in the Standard Model,
  the factors of $\frac{5}{3}$ in eqs. \eqref{eq:6}, \eqref{eq:8} and
  \eqref{eq:10} have to be omitted, while for eq. \eqref{eq:22} one
  should use $b_1=11$.}  Similarly, the constraints on
$\delta\alpha/\alpha$ in Table~\ref{tab:constraints} and the allowed
range for $R$ in eq. \eqref{eq:26} hold in these more general
circumstances.

An important comment is now in order: In principle, the variation
$\delta\mu_{\rm th}$ may induce a variation in all couplings at
lower scales. In the most general case, this includes a
variation in the masses of all the particles which are lighter than
$\mu_{\rm th}$. A variation of these masses
could be induced by two sources:
\begin{enumerate}
\item The parameters that determine a particle mass, such as Yukawa
  couplings, may have threshold corrections at $\mu_{\rm th}$. In
  this case, a variation analogous to~\eqref{eq:4} results. To
  calculate this effect, one should use the (linearized) beta function
  of these parameters below and above the scale $\mu_{\rm th}$ .
\item The variation in the gauge couplings, $\alpha_i$, below
  $\mu_{\rm th}$, leads to variations in the low scale masses. To
  calculate this effect, one should use the RGE of the mass
  parameters below $\mu_{\rm th}$.
\end{enumerate}
When running the gauge couplings from $\mu_{\rm th}$ to some low scale
$Q$, all particles with masses $Q<m<\mu_{\rm th}$ are integrated out. For
example, in an MSSM GUT model, with $Q\sim m_b$ and  $\mu_{\rm
  th}=M_{\rm GUT}$, we integrate out all supersymmetric particles at a
scale $M_{\rm SUSY}$ as well as heavy SM particles ($t,Z,W^\pm$). This
integration out would usually introduce additional threshold
corrections, which have small effects on the parameters at $m_b$.

We are mainly interested in small variations of $\alpha$ and
$\lqcd$. As explained above, the variation $\delta\mu_{\rm th}$
induces variations in all intermediate scales below $\mu_{\rm
  th}$. But these variations induce, by threshold corrections
at these scales, further variation in gauge couplings.
The whole process may be pictured as a `chain reaction', with a final
effect that could be significant, even if the individual variations
at each intermediate scale are small. We postpone the detailed study
of this effect to future work.

Here, we would only like to emphasize that when $R$ is large, as is
the case in most of the GUT models that we have investigated, the
variation of $\lqcd$ is the dominant effect at low energies, and it is
well justified to ignore variations in couplings other than the gauge
couplings. For small $R$, however, the modification due to threshold
corrections of other parameters can become important. Thus, when we
think of $R=0$ models, one should not conclude that $\alpha$ would
vary with $\lqcd$ remaining strictly fixed. Very likely, $\lqcd$
varies due to the effects of lower thresholds and of higher loops. But the
effect is expected to be small [$R={\cal O}(1)$] and our qualitative
conclusions would not change.

\section{Conclusions}
\label{sec:conclusions}
A variation in a physical scale in which threshold corrections take
place leads to variations in low scale observables. Current
experimental data constrain the relations between the threshold
corrections of the three different gauge couplings. In particular,
consistency between the claimed variation $\delta\alpha/\alpha$
[eq. \eqref{tivaal}] and the allowed ranges for $\delta X/X$ and
$\delta Y/Y$ [Table~\ref{tab:constraints}] give
\begin{equation}\label{eq:final}
-1\lesssim R\equiv{2\pi\over9\alpha}{\Delta b_3\over\frac{5}{3}\Delta
  b_1+\Delta b_2}\lesssim +6,
\end{equation}
where $\Delta b_i$ is the difference of the one-loop beta function
coefficient above and below the varying scale. In other words,
the difference between the beta function coefficients
of the strong coupling $\alpha_3$ above and below the varying scale,
should be suppressed by a factor of $\mathcal{O}(0.01)$ compared with
those of the other two couplings, $\alpha_1$ and $\alpha_2$.

We focussed our investigation on the framework where the MSSM is
embedded in GUT models and the varying scale is $M_{\rm GUT}$. We
demonstrated that the relation between the variation of $\lqcd$ and
that of $\alpha$, that is, the ratio
$R=\frac{\delta\lqcd/\lqcd}{\delta\alpha/\alpha}$, is highly model
dependent. For example, we can construct SU(5) models with $R$ as high
as ${\cal O}(+600)$ and as low as ${\cal O}(-500)$, and a model where
$R=0$. (The latter is the minimal SU(5) model, and the former have,
respectively, eight and seven ${\bf5}+\overline{\bf5}$ pairs added to
the minimal model.) We argued, however, that it is difficult to obtain
consistency with the requirement \eqref{eq:final} in realistic
supersymmetric GUT models in our framework.

\begin{acknowledgments}
The visit of M.D. to the Weizmann Institute was supported by the Albert
Einstein Minerva Center for Theoretical Physics and by the United
States - Israel Binational Science Foundation (BSF).
Y.N.\ is supported by the Israel Science Foundation founded by the
Israel Academy of Sciences and Humanities.
\end{acknowledgments}



\end{document}